\documentstyle[12pt,fleqn]{article}
\def\thebibliography#1{\section*{References}\list
  {[\arabic{enumi}]}{\settowidth\labelwidth{#1}\leftmargin\labelwidth
    \advance\leftmargin\labelsep
    \usecounter{enumi}}
    \def\newblock{\hskip .11em plus .33em minus .07em}
    \sloppy\clubpenalty4000\widowpenalty4000
    \sfcode`\.=1000\relax}

\let\Large=\large

\def\op#1{\mathop{{\it\fam0} #1}\limits}

\newcommand{\Ker}{{\rm Ker\,}}

\newcommand{\beq}{\begin{equation}}
\newcommand{\eeq}{\end{equation}}
\newcommand{\ben}{\begin{eqnarray}}
\newcommand{\een}{\end{eqnarray}}
\newcommand{\be}{\begin{eqnarray*}}
\newcommand{\ee}{\end{eqnarray*}}
\newcommand{\bea}{\begin{eqalph}}
\newcommand{\eea}{\end{eqalph}}

\newcommand{\bR}{{\bf R}}

\newcommand{\rrq}{{\ol q}}

\newcommand{\al}{\alpha}
\newcommand{\bt}{\beta}

\newcommand{\la}{\lambda}

\newcommand{\f}{\phi}

\newcommand{\m}{\mu}

\newcommand{\g}{\gamma}
\newcommand{\G}{\Gamma}

\newcommand{\si}{\sigma}

\newcommand{\wt}{\widetilde}

\newcommand{\ol}{\overline}

\newcommand{\dr}{\partial}

\newcommand{\ar}{\op\longrightarrow}

\newcommand{\ot}{\otimes}

\newcounter{eqalph}
\newcounter{equationa}
\newcounter{remark}
\newcounter{theorem}
\newcounter{proposition}
\newcounter{lemma}
\newcounter{corollary}
\newcounter{definition}

\newenvironment{eqalph}{\stepcounter{equation}
\setcounter{equationa}{\value{equation}}
\setcounter{equation}{0}

\begin{eqnarray}}{\end{eqnarray}\setcounter{equation}{\value{equationa}}}

\def\theremark{\arabic{remark}}

\def\thedefinition{\arabic{definition}}

\newenvironment{rem}{\refstepcounter{remark}\bigskip\noindent{\it
Remark \theremark.}}{\medskip}

\newenvironment{prop}{\refstepcounter{definition} 
\bigskip\noindent{\it Proposition \thedefinition.}}{\medskip}
\newenvironment{lem}{\refstepcounter{definition} 
\bigskip\noindent{\it Lemma \thedefinition.}}{\medskip}
\newenvironment{cor}{\refstepcounter{definition} 
\bigskip\noindent{\it Corollary \thedefinition.}}{\medskip}

\begin{document}

\hbox{}

{\parindent=0pt 

{ \Large \bf On the geodesic form of non-relativistic dynamic equations}
\bigskip

{\sc Luigi Mangiarotti$\dagger$\footnote{E-mail
address: mangiaro@camserv.unicam.it} and Gennadi 
Sardanashvily$\ddagger$\footnote{E-mail address: sard@grav.phys.msu.su}}
\medskip

\begin{small}
$\dagger$ Department of Mathematics and Physics, University of Camerino, 62032
Camerino (MC), Italy \\
$\ddagger$ Department of Theoretical Physics, Physics Faculty, Moscow State
University, 117234 Moscow, Russia
\bigskip

{\bf Abstract.} It is shown
that any second order dynamic equation on a configuration bundle $Q\to\bR$
of non-relativistic mechanics 
is equivalent to  a geodesic equation with respect to a
(non-linear) connection on the tangent bundle
$TQ\to Q$. The case of quadratic dynamic equations is analyzed in details.
The equation for Jacobi vector fields is
constructed and investigated by the 
geometric methods. 
\end{small}
}

\section{Introduction}

We are concerned with non-relativistic mechanics on a
configuration bundle $Q\to\bR$, where $\bR$ is the time axis. The corresponding
velocity phase space is the first order jet manifold $J^1Q$ of sections of
$Q\to\bR$. A second order dynamic equation
(called further simply a dynamic equation) on a fibre bundle $Q\to\bR$ is
defined as a first order dynamic equation on the jet bundle $J^1Q\to \bR$,
given by a holonomic connection
$\xi$ on $J^1Q\to
\bR$ which takes its values in the second order jet manifold $J^2Q\subset
J^1QJ^1Q$ (see, e.g., \cite{krupva,leon,book98,massa}). This connection
$\xi$ is also called a semispray vector field \cite{leon}, a SODE field
\cite{cramp}, a special vector field \cite{giach92} because of the canonical
imbedding $J^1J^1Q\to TJ^1Q$. 

The fact that $\xi$ is a flat connection places
a limit on the geometric analysis of non-relativistic dynamic equations.
Nevertheless, it  was proved that  
every dynamic equation $\xi$  defines a connection $\g$ on the affine jet
bundle
$J^1Q\to Q$, and {\it vice versa} \cite{cramp,giach92,leon,book98}. For the
sake of simplicity, we call
$\g$ a dynamic connection, but this is not the terminology of \cite{massa},
where this term stands for a linear connection on the tangent bundle $TJ^1Q\to
J^1Q$ associated with each dynamic equation $\xi$ too 
\cite{cramp,giach92,book98}). 
Here, we show that, due to the canonical imbedding
$J^1Q\to TQ$, every dynamic connection  
yields a
(non-linear) connection on the tangent bundle
$TQ\to Q$, and {\it vice versa}. As a consequence, every dynamic equation on
$Q$  gives rise to an equivalent geodesic equation on the tangent bundle
$TQ\to Q$ in accordance with the following Proposition.

\begin{prop} \label{c1}  Given a configuration bundle $Q\to\bR$
coordinated by $(q^0=t,q^i)$ and  its second order jet manifold $J^2Q$ 
coordinated by
$(q^\la,q^i_t,q^i_{tt})$, any dynamic equation 
\beq
q^i_{tt} =\xi^i(t,q^j,q_t^j) \label{cqg5}
\eeq
of non-relativistic mechanics on $Q\to\bR$ is equivalent to the geodesic
equation 
\ben
 && \ddot q^0 =0, \qquad \dot q^0=1, \nonumber\\
 && \ddot q^i= \wt K^i_0 + \wt K^i_j\dot q^j
\label{cqg11}
\een
with respect to a connection
$\wt K$ on $TQ\to Q$ which fulfills the conditions
\beq
\wt K^0_\la =0, \qquad
 \xi^i= \wt K^i_0 + q^j_t\wt K^i_j\mid_{\dot q^0=1,\dot q^i=q^i_t}.
\label{cqg9}
\eeq 
\end{prop}

\begin{rem}
Recall that, in conservative mechanics, a
second order dynamic equation on a configuration manifold $M$ is defined as a
particular holonomic vector field $\Xi$ on the tangent bundle $TM$. This
dynamic equation yields a connection on $TM\to M$, but fails to be a geodesic
equation in general \cite{mora}. Nevertheless, every second order dynamic
equation on $M$ gives rise to a  dynamic equation on the fibre
bundle
$\bR\times M\to\bR$ (see Remark \ref{cons}), and can be written as a geodesic
equation in accordance with Proposition \ref{c1}.
\end{rem}

Since a configuration bundle $Q\to\bR$ is trivial, the existent
formulations of non-relativistic mechanics often imply its preliminary
splitting $Q=\bR\times M$
\cite{cari93,eche,leon,mora}. 
This is not the case of mechanical systems
subject to time-dependent transformations, including inertial frame
transformations. Recall that different trivializations of $Q\to\bR$ differ from
each other in projections $Q\to M$. 
Since a configuration bundle $Q\to\bR$ has no
canonical trivialization in general, mechanics on $Q\to\bR$
is not a repetition of mechanics on $\bR\times M$, but implies additionally
a connection on $Q\to\bR$ which is a non-relativistic reference frame
\cite{book98,sard98} (see, e.g., Proposition \ref{jp51}).
Proposition \ref{c1} shows that, considered independently on a trivialization
of $Q\to\bR$,  non-relativistic dynamic equations make the geometric sense of
geodesic equations. Treated in such a way, non-relativistic dynamic equations
can be examined by means of the differential geometric methods. For instance,
the curvature of the connection $\wt K$ is called into play (see Propositions
\ref{nl5},
\ref{1}, \ref{2}). 

Using Proposition \ref{c1}, we examine quadratic dynamic equations in details.
In this case, the corresponding dynamic connection
$\g$ on
$J^1Q\to Q$ is affine, while the connection $\wt K$ (\ref{cqg9}) on $TQ\to Q$ is
linear. Then the equation for Jacobi vector fields along the geodesics of the
connection
$\wt K$ can be considered. This equation coincides with
the existent equation for Jacobi fields of a Lagrangian system
\cite{ditt,book98} in the case of non-degenerate quadratic Lagrangians, when
they can be compared. We will consider more general case
of quadratic Newtonian systems characterized by a
pair $(\xi,m)$ of a quadratic dynamic equation $\xi$ and a Riemannian mass
tensor
$m$ which satisfy a certain compatibility condition. Given a reference frame,
a Riemannian mass tensor
$m$ is  extended to a Riemannian metric on the configuration space
$Q$. Then conjugate points of solutions of the dynamic equation $\xi$ can be
examined in accordance with the well-known geometric criteria.

\section{Technical preliminaries}

A configuration bundle $Q\to\bR$ of non-relativistic mechanics throughout is
coordinated by
$(t,q^i)$, where $t$ is a Cartesian coordinate on the
time axis
$\bR$ with the transition functions
$t'=t+$const. We will use the compact notation 
$(q^{\la=0}=t, q^i)$, 
$\dr_\la=\dr/\dr q^\la$, $\dot \dr_\la=\dr/\dr \dot q^\la$.  
The velocity phase space $J^1Q$ 
is provided with the adapted coordinates
$(q^\la,q^i_t)$. 

Recall that the first order jet manifold $J^1Q$ comprises the equivalence
classes
$j^1_tc$ of sections of
$Q\to\bR$ which are identified by their values  $c^i(t)$ and the values of
their partial derivatives
$\dr_tc^i(t)$ at points $t\in\bR$, i.e., $q^i_t(j^1c)= \dr_tc^i(t)$ (see,
e.g., \cite{kol,book,book99,sau}). There is the canonical imbedding 
\beq
\la_1: J^1Q\op\hookrightarrow_Q TQ, \qquad \la_1=d_t=\dr_t
+q^i_t\dr_i,\label{z260}
\eeq
where $d_t$ denotes the total derivative. 
From now on, we will 
identify $J^1Q$ with its image in $TQ$. 
This is an
affine bundle modelled over the vertical tangent bundle $VQ$ of
$Q\to\bR$. 

As a consequence of (\ref{z260}), every connection 
\beq
\G:Q\to J^1Q, \qquad \G=dt\ot(\dr_t +\G^i\dr_i), \label{z270}
\eeq
on a fibre bundle $Q\to\bR$ is identified with the
nowhere  vanishing vector field 
\beq
\G:Q\to J^1Q\subset TQ, \qquad \G=\dr_t +\G^i\dr_i, \label{1005}
\eeq
on $Q$ \cite{book98,book99}. 
This is the horizontal lift of the
standard vector field $\dr_t$ on $\bR$ by means of the connection 
(\ref{z270}). Conversely, any vector field $\G$ on $Q$ such that
$dt\rfloor\G =1$ defines a connection on $Q\to\bR$.
Accordingly, the covariant differential associated with a
connection $\G$ on $Q\to\bR$ reads
\beq
 D^\G: J^1Q\op\to_Q VQ, \qquad 
\dot q^i\circ D_\G =q^i_t-\G^i. \label{z279} 
\eeq

By $J^1J^1Q$ is meant the first order jet manifold of the jet bundle
$J^1Q\to\bR$, coordinated by  $(q^\la,q^i_t, q^i_{(t)},q^i_{tt})$. 
The second order jet manifold $J^2Q$ of the fibre bundle $Q\to\bR$ is
the holonomic subbundle $q^i_t= q^i_{(t)}$ of $J^1J^1Q$, coordinated by
$(q^\la,q^i_t,q^i_{tt})$. 
There are the imbeddings
\ben
&& J^2Q \ar^{\la_2} TJ^1Q \ar^{T\la_1} T^2Q, \nonumber
\\ && \la_2: (q^\la,q^i_t,q^i_{tt})\mapsto (q^\la,q^i_t,\dot
q^0=1,\dot q^i=q^i_t,\dot q^i_t=q^i_{tt}). \label{gm211} \\
&& T\la_1\circ \la_2:(q^\la,q^i_t,q^i_{tt})\mapsto
 (q^\la, \dot q^0 =1,
\dot q^i =q^i_t, \ddot q^0=0, \ddot q^i=q^i_{tt}),
\label{cqg80}
\een 
where $(q^\la,\dot q^\la, \ddot
q^\la)$ are holonomic coordinates on 
$T^2Q$. 
By $J^1_QJ^1Q$ is meant the first order jet manifold of
the affine jet bundle
$J^1Q\to Q$. The adapted coordinates on $J^1_QJ^1Q$ are
$(q^\la,q^i_t,q^i_{\la t})$.  

\section{Geometry of non-relativistic mechanics}

This Section is devoted to the proof of Proposition 1.

As was mentioned above, a dynamic equation on a
configuration bundle
$Q\to\bR$ is defined as the geodesic equation $\Ker D^\xi\subset J^2Q$
for a holonomic connection $\xi$ on the jet bundle $J^1Q\to\bR$.
 It is given by the coordinate expression (\ref{cqg5}).
Due to the morphism (\ref{gm211}), a holonomic connection $\xi$  is
represented by the horizontal vector field on $J^1Q$
\beq
\xi=\dr_t + q^i_t\dr_i + \xi^i(q^\m,q^i_t) \dr_i^t. \label{a1.30}
\eeq

\begin{rem} \label{cons} 
A dynamic equation $\xi$ is said to be conservative if
there exists a trivialization $Q\cong \bR\times M$ such that the vector field 
$\xi$ (\ref{a1.30}) on $J^1Q\cong \bR\times TM$ is projectable onto $TM$. Then
this projection
\beq
\Xi_\xi=\dot q^i\dr_i +\xi^i(q^j,\dot q^j)\dot \dr_i \label{jp26}
\eeq
is a second order dynamic equation
\beq
\ddot q^i=\Xi_\xi^i \label{nl1}
\eeq
on the typical fibre $M$ of $Q\to\bR$. Conversely, every
second order dynamic equation $\Xi$ (\ref{nl1}) on a manifold $M$ can be seen
as a conservative dynamic equation
\beq
\xi_\Xi=\dr_t + \dot q^i\dr_i + u^i\dot \dr_i \label{jp27}
\eeq
on the fibre bundle $\bR\times M\to\bR$. 
\end{rem}

Let us turn to the above mentioned relationship
between the holonomic connections $\xi$ (\ref{a1.30}) on $J^1Q\to\bR$ and the
dynamic connections 
\beq
 \g=dq^\la\ot (\dr_\la + \g^i_\la \dr_i^t)
\label{a1.38} 
\eeq 
on the affine jet bundle $J^1Q\to Q$ (see, e.g., \cite{book98}). 

\begin{prop}\label{gena51} 
Any dynamic connection $\g$ (\ref{a1.38}) defines
the holonomic connection 
\beq
\xi_\g = \dr_t + q^i_t\dr_i +(\g^i_0 +q^j_t\g^i_j)\dr_i^t
\label{z281}
\eeq
on $J^1Q\to\bR$.
Conversely, any holonomic connection $\xi$ (\ref{a1.30}) on 
$J^1Q\to \bR$ defines the dynamic connection 
\beq
\g_\xi =dt\ot[\dr_t+(\xi^i-\frac12 q^j_t\dr_j^t\xi^i)\dr_i^t] +
dq^j\ot[\dr_j +\frac12\dr_j^t\xi^i \dr_i^t].
\label{z286}
\eeq
\end{prop}

It follows that every dynamic connection $\g$ (\ref{a1.38}) yields the
non-relativistic dynamic equation 
\beq
q^i_{tt}=\g^i_0 +q^j_t\g^i_j \label{z287}
\eeq
(\ref{cqg5}) on the configuration bundle  $Q\to \bR$. 
Different dynamic connections may lead to the same dynamic equation
 (\ref{z287}). 
The dynamic connection $\g_\xi$ (\ref{z286}),
associated with a dynamic equation,  possesses the property
\be
\g^k_i = \dr_i^t\g^k_0 +  q^j_t\dr_i^t\g^k_j,
\ee
which implies the relation $\dr_j^t\g^k_i = \dr_i^t\g^k_j$. Such
a dynamic connection is
called symmetric.
Let $\g$ be a dynamic connection (\ref{a1.38}) and $\xi_\g$ the 
corresponding dynamic equation (\ref{z281}). Then 
the connection (\ref{z286}), associated with $\xi_\g$, takes the form
\be
\g_{\xi_\g}{}^k_i = \frac{1}{2}
(\g^k_i + \dr_i^t\g^k_0 + q^j_t\dr_i^t\g^k_j),
\qquad \g_{\xi_\g}{}^k_0 = \xi^k - q^i_t\g_{\xi_\g}{}^k_i. 
\ee
It is readily observed that $\g = \g_{\xi_\g}$ if and only if $\g$ is
symmetric.
 
Now let us turn to the proof of Proposition \ref{c1}.

We start from the relation between the connections $\g$ (\ref{a1.38}) on the
affine jet bundle $J^1Q\to Q$ and the connections
\beq
K=dq^\la\ot(\dr_\la +K_\la^\al\dot \dr_\al) \label{z290}
\eeq 
on the tangent bundle $TQ\to Q$.  Let us
consider the diagram
\beq
\begin{array}{rcccl}
& J^1_QJ^1Q & \ar^{J^1\la_1} & J^1_QTQ & \\
_\g &  \put(0,-10){\vector(0,1){20}} & &  \put(0,-10){\vector(0,1){20}}
& _K\\
& J^1Q &\ar^{\la_1} & TQ &
\end{array} \label{z291}
\eeq
where $J^1_QTQ$ is the first order jet manifold of the tangent bundle $TQ\to
Q$, coordinated by $(q^\la,\dot q^\la,\dot q^\la_\m)$.
The jet prolongation over $Q$ of the canonical imbedding $\la_1$
(\ref{z260}) reads 
\be
J^1\la_1: (q^\la,q^i_t, q^i_{\m t}) \mapsto 
(q^\la,\dot q^0=1,\dot q^i=q^i_t, \dot q_\m^0=0,
\dot q^i_\m=q^i_{\m t}).
\ee
We have
\be
&& J^1\la_1\circ \g: (q^\la,q^i_t) \mapsto 
(q^\la,\dot q^0=1,\dot q^i=q^i_t, \dot q_\m^0=0,
\dot q^i_\m=\g^i_\m ),\\
&& K\circ \la_1: (q^\la,q^i_t) \mapsto 
(q^\la,\dot q^0=1,\dot q^i=q^i_0, \dot q^0_\m=K_\m^0,
\dot q^i_\m=K^i_\m).
\ee
It follows that the diagram (\ref{z291}) can be commutative only
if the components $K^0_\m$ of the connection $K$ on $TQ\to
Q$ vanish. 
Since the transition functions $t\to t'$ are independent of
$q^i$, a connection 
\beq
\wt K=dq^\la\ot (\dr_\la +K^i_\la\dot\dr_i)
\label{z292}
\eeq
with the components $K^0_\m=0$ can exist on the tangent
bundle
$TQ\to Q$. It 
obeys the transformation law
\beq
{K'}_\la^i=(\dr_j x'^i K^j_\m + \dr_\m\dot x'^i)
\frac{\dr q^\m}{\dr x'^\la}.  \label{z293}
\eeq
Now the diagram (\ref{z291}) becomes commutative if the connections
$\g$ and $\wt K$ fulfill the relation
\beq
\g^i_\m=K^i_\m(q^\la,\dot q^0=1, \dot q^i=q^i_t), \label{z294}
\eeq
which holds globally since the substitution of $\dot q^i= q^i_t$
into (\ref{z293}) restates the coordinate transformation law of 
$\g$.  In accordance with this relation, a desired connection
$\wt K$ is an extension  of the local section
$J^1\la_1\circ \g$ of the affine bundle $J^1_QTQ\to TQ$ over the closed
submanifold $J^1Q\subset TQ$ to a global section. Such an extension
always exists, but is not unique. Thus, it is stated the following.

\begin{prop}\label{mot1} 
Every non-relativistic dynamic equation (\ref{cqg5}) on the configuration
bundle
$Q\to\bR$ can be written in the form 
\beq
q^i_{tt} = K^i_0\circ\la_1 +q^j_t K^i_j\circ\la_1, \label{gm340}
\eeq
where $\wt K$ is a connection (\ref{z292}) on the tangent bundle $TQ\to Q$.
Conversely, each connection $\wt K$ (\ref{z292}) on $TQ\to
Q$ defines the dynamic equation
(\ref{gm340}) on $Q\to\bR$.
\end{prop}

Let us consider the geodesic equation (\ref{cqg11}) on $TQ$ with respect to
the connection $\wt K$. 
Its solution is a geodesic curve $c(t)$ which also
satisfies the dynamic equation (\ref{cqg5}), and {\it vice versa}. It states
 Proposition \ref{c1}.

\section{Non-relativistic reference frames}

Proposition \ref{mot1} gives more than it is needed for Proposition
\ref{c1}, and we can prove a converse of Proposition \ref{c1}.

Let us start from the notion of a reference frame in non-relativistic
mechanics. From the physical viewpoint,
a reference frame in non-relativistic mechanics on a configuration
bundle $Q\to\bR$ sets a tangent vector at each point of
$Q$ which characterizes the velocity of an "observer" at this point. Then 
any connection
$\G$ on $Q\to\bR$ is said to be such a
reference frame \cite{eche95,book98,massa,sard98}.

\begin{lem}\label{gena113} \cite{book98,book99}.
 Each connection $\G$ on
a fibre bundle $Q\to\bR$ defines an atlas of local constant trivializations of
$Q\to\bR$ 
whose transition
functions are independent of $t$, and {\it vice versa}.
One finds $\G=\dr_t$ 
with respect to this atlas.  
In particular, there is one-to-one correspondence between the complete
connections
$\G$ (\ref{1005}) on
$Q\to\bR$ and the
trivializations of this bundle.
\end{lem}

By virtue of this Lemma, any  coordinate atlas $(t, q^i)$ on $Q\to\bR$ whose
transition functions are independent of time is also regarded as a
reference frame.
Using the notion of a reference frame, we can formulate a desired converse of
Proposition \ref{c1}.

\begin{prop}\label{jp51} 
Given a reference frame $\G$, any connection $K$
(\ref{z290}) on the tangent bundle $TQ\to Q$ defines the dynamic equation 
\beq
\xi^i= (K^i_\la -\G^i K^0_\la)\dot q^\la\mid_{\dot q^0=1,\dot q^i=q^i_t}.
\label{jp52}
\eeq
\end{prop}

The proof follows at once from Proposition \ref{mot1} and the following
assertion.

\begin{lem} 
Given a connection $\G$ on the fibre bundle $Q\to\bR$ and a connection $K$ on
the tangent bundle $TQ\to Q$, there is the connection $\wt K$ on $TQ\to Q$ with
the components
\be
\wt K^0_\la =0, \qquad \wt K^i_\la = K^i_\la - \G^iK^0_\la.
\ee
It is proved by the inspection of transition functions.
\end{lem}

\section{Quadratic dynamic equations}

From the physical viewpoint, the most interesting dynamic equations are the
quadratic ones, i.e.,
\beq
\xi^i = a^i_{jk}(q^\m)q^j_t q^k_t + b^i_j(q^\m)q^j_t + f^i(q^\m).
\label{cqg100}
\eeq
This property is coordinate-independent due to the affine transformation law
of coordinates $q^i_t$. Then, it is readily observed that the corresponding
dynamic connection
$\g_\xi$ (\ref{z286}) is affine:
\be
 \g=dq^\la\ot [\dr_\la + (\g^i_{\la 0}(q^\nu)+ \g^i_{\la
j}(q^\nu)q^j_t)\dr_i^t],
\ee
 and {\it vice versa}.
This connection is symmetric if and only if $\g^i_{\la \m}=\g^i_{\m\la}$. 

\begin{lem}\label{aff}
There is one-to-one correspondence between the affine connections $\g$ on
the affine jet bundle $J^1Q\to Q$ and the linear connections $\wt K$
(\ref{z292}) on the tangent bundle $TQ\to Q$. 
\end{lem}

This correspondence is
given by the relation (\ref{z294}) which takes the form
\be
\g^i_\m=\g^i_{\m 0} + \g^i_{\m j}q^j_t,  \g^i_{\m\la}= K_\m{}^i{}_\la. 
\ee
In particular, if an affine dynamic connection $\g$ is symmetric, so is the
corresponding linear connection $\wt K$.

Then we come to the following corollaries of Propositions \ref{c1},
\ref{jp51}.

\begin{cor}\label{c2} 
Any quadratic dynamic equation
\beq
q^i_{tt}= a^i_{jk}(q^\m)q^j_t q^k_t + b^i_j(q^\m)q^j_t + f^i(q^\m)
\label{cqg100'}
\eeq
is equivalent to the geodesic
equation 
\ben
&& \ddot q^0= 0, \qquad \dot q^0=1,\nonumber\\
&& \ddot q^i= 
a^i_{jk}(q^\m)\dot q^i \dot q^j + b^i_j(q^\m)\dot q^j\dot q^0 +
f^i(q^\m) \dot q^0\dot q^0. \label{cqg17}
\een
for the symmetric linear connection 
\be
\wt K=dq^\la\ot(\dr_\la + K_\la{}^\m{}_\nu(q^\al)\dot q^\nu\dot\dr_\m)
\ee
on $TQ\to Q$,
given by the components
\beq
K_\la{}^0{}_\nu=0, \quad K_0{}^i{}_0= f^i, \quad
K_0{}^i{}_j=K_j{}^i{}_0=\frac12 b^i_j,
\quad K_j{}^i{}_k= a^i_{jk}. \label{cqg101}
\eeq
\end{cor}

\begin{cor}\label{aff1}
Conversely, any linear connection $K$  on the tangent bundle
$TQ\to Q$ defines the quadratic dynamic equation
\be
q^i_{tt}= K_j{}^i{}_kq^j_tq^k_t + (K_0{}^i{}_j+K_j{}^i{}_0)q^j_t +
K_0{}^i{}_0,
\ee
written with respect to a given reference frame $(t,q^i)$.
\end{cor}

The geodesic equation
(\ref{cqg17}) however is not unique for the dynamic equation (\ref{cqg100'}).

\begin{prop} \label{jp40} 
Any quadratic dynamic equation (\ref{cqg100}), being 
equivalent to the geodesic equation with respect to the linear connection
$\wt K$ (\ref{cqg101}), is also equivalent to the geodesic equation with
respect to an affine connection $K'$ on $TQ\to Q$ which differs from $\wt K$
(\ref{cqg101}) in a soldering form $\si$ on $TQ\to Q$ with the components
\be
\si^0_\la= 0, \qquad \si^i_k= h^i_k+ (s-1) h^i_k\dot q^0, \qquad \si^i_0=
-s h^i_k\dot q^k -h^i_0\dot q^0 + h^i_0,
\ee
where $s$ and $h^i_\la$ are local functions on $Q$.
\end{prop} 

\section{Free motion equation}

Let us point out the following interesting class of dynamic equations which
we agree to call the free motion equations.   

We say that the dynamic equation (\ref{cqg5}) is a free motion
equation if there exists a reference frame $(t,\ol q^i)$ on the configuration
bundle $Q\to\bR$ 
 such that this equation reads 
\beq
\ol q^i_{tt}=0. \label{z280}
\eeq
With respect to arbitrary bundle coordinates
$(t,q^i)$, a free motion equation takes the form
\beq
 q^i_{tt}=d_t\G^i +\dr_j\G^i(q^j_t-\G^j) -
\frac{\dr q^i}{\dr\rrq^m}\frac{\dr\rrq^m}{\dr q^j\dr q^k}(q^j_t-\G^j)
(q^k_t-\G^k),  \label{m188}
\eeq
where $\G^i=\dr_t q^i(t,\ol q^j)$ is the connection associated with the initial
frame $(t,\ol q^i)$.  One can think of the right hand side of the equation
(\ref{m188}) as being the general coordinate expression of an inertial force
in non-relativistic mechanics. The corresponding
dynamic connection $\g$ on the affine jet bundle $J^1Q\to Q$ reads
\beq
 \g^i_k=\dr_k\G^i  -
\frac{\dr q^i}{\dr\rrq^m}\frac{\dr\rrq^m}{\dr q^j\dr q^k}(q^j_t-\G^j),
\qquad
\g^i_0= \dr_t\G^i +\dr_j\G^iq^j_t -\g^i_k\G^k. \label{gm330}
\eeq
It is affine. By virtue of Lemma \ref{aff}, this dynamic connection
defines a linear connection $K$ on the tangent bundle $TQ\to Q$ whose
curvature is necessarily equal to 0.  Thus, we come to the following
criterion of a dynamic equation to be a free motion equation.

\begin{prop} \label{nl5} 
If $\xi$ is a free motion
equation, it is quadratic and the corresponding
linear symmetric connection (\ref{cqg101}) on the tangent bundle $TQ\to Q$ is
flat.
\end{prop}

This criterion fails to be a sufficient condition since it may happen that the
components of a curvature-free linear symmetric connection on $TQ\to Q$ vanish
with respect to the coordinates on $Q$
 which are not compatible with the fibration $Q\to\bR$. Nevertheless, one can
formulate the necessary and sufficient condition of the existence of a free
motion equation on a configuration space $Q$.

\begin{prop} \cite{cramp,book98}.
A free motion equation on an a configuration bundle $Q\to\bR$ exists if and
only if the typical fibre $M$ of $Q$ admits a curvature-free linear symmetric
connection.
\end{prop}

\section{Quadratic Lagrangian and Newtonian systems}

A Lagrangian of a mechanical system on $Q\to\bR$ is defined as a function on 
the velocity phase space $J^1Q$.
Let us consider a non-degenerate quadratic Lagrangian 
\beq
L=\frac12m_{ij}(q^\m) q^i_t q^j_t + k_i(q^\m) q^i_t  +
f(q^\m), \label{cqg20}
\eeq
where $m_{ij}$ is a Riemannian fibre metric in the vertical tangent bundle
$VQ$, called a mass metric. As for quadratic dynamic equations, this property
is coordinate-independent. Similarly to Lemma
\ref{aff}, one can show that any quadratic polynomial on $J^1Q\subset TQ$ is
extended to a bilinear form on $TQ$. Then the Lagrangian $L$
(\ref{cqg20}) can be written as 
\be
L=\frac12g_{\al\m}q^\al_t q^\m_t, \qquad q^0_t=1,
\ee
where $g$ is the (degenerate) fibre metric
\beq
g_{00}=2f, \qquad g_{0i}=k_i, \qquad g_{ij}=m_{ij} \label{cqg21}
\eeq
in the tangent bundle $TQ$. The associated Lagrange equation takes the form
\beq
q^i_{tt}=(m^{-1})^{ik}\{_{\la k\nu}\}q^\la_tq^\nu_t, \qquad q^0_t=1,
\label{cqg35}
\eeq
where 
\be
\{_{\la\m\nu}\} =-\frac12(\dr_\la g_{\m\nu} +\dr_\nu
g_{\m\la} - \dr_\m g_{\la\nu})
\ee
 are the Christoffel symbols of the metric (\ref{cqg21}). The corresponding
geodesic equation (\ref{cqg17}) on
$TQ$  reads
\ben
&& \ddot q^0 = 0, \qquad \dot q^0=1, \nonumber \\
&&\ddot q^i = (m^{-1})^{ik}\{_{\la k\nu}\}\dot q^\la\dot q^\nu, \label{cqg47}
\een
where $\wt K$ (\ref{cqg9}) is a linear connection with the components
\beq
\wt K_\la{}^0{}_\nu=0, \qquad \wt K_\la{}^i{}_\nu=(m^{-1})^{ik}\{_{\la k\nu}\}.
\label{nl6}
\eeq
We have the relation 
\beq
\dot q^\la (\dr_\la m_{ij} + K_\la{}^i{}_\nu \dot q^\nu)=0. \label{nl7}
\eeq

One can show that an arbitrary Lagrangian system on a configuration bundle
$Q\to\bR$ is a particular Newtonian system on $Q\to\bR$. The latter is
defined as a pair $(\xi,m)$ of a dynamic equation $\xi$ and a
(degenerate) fibre metric $m$ in the fibre bundle $V_QJ^1Q\to J^1Q$ which
satisfy the symmetry condition $\dr_k^tm_{ij}=\dr^t_j m_{ik}$ and the
compatibility condition 
\beq
\xi\rfloor dm_{ij} + m_{ik}\g^k_j + m_{jk}\g^k_i = 0, \label{a1.95}
\eeq
where $\g_\xi$ is the dynamic connection (\ref{z286}) \cite{book98,book99}.
Note that the compatibility condition (\ref{a1.95}) can be written in an
intrinsic way as
$\ol \nabla_\xi m=0$, where $\ol\nabla$ is the covariant derivative with
respect to the canonical prolongation of the connection $\g_\xi$ onto the
vertical cotangent bundle $J^*_QJ^Q\to J^1Q$. 

We will restrict our consideration to non-degenerate quadratic Newtonian
systems when
$\xi$ is a quadratic dynamic equation (\ref{cqg100}) and $m$ is a Riemannian
fibre metric in
$VQ$, i.e., $m$ is independent of $q^i_t$ and the symmetry condition becomes
trivial. In this case, the dynamic equation (\ref{cqg100'}) is equivalent to
the geodesic equation (\ref{cqg17}) with respect a symmetric linear connection
$\wt K$ (\ref{cqg101}), while the compatibility condition (\ref{a1.95}) takes
the form (\ref{nl7}). 

Given a symmetric linear connection $\wt K$ (\ref{cqg101}) on the tangent
bundle
$TQ\to Q$, one can consider the equation for Jacobi vector fields along
geodesics of this connection, i.e., along solutions of the non-relativistic
dynamic equation (\ref{cqg100'}). If $Q$ is provided with a Riemannian
metric, the conjugate points of these geodesic can be investigated.

\section{Non-relativistic Jacobi fields}

Let us consider the quadratic dynamic equation (\ref{cqg100'}) and the
equivalent geodesic equation (\ref{cqg17}) with respect to the symmetric
linear connection $\wt K$ (\ref{cqg101}). Its curvature
\be
R_{\la\m}{}^\al{}_\bt =\dr_\la K_\m{}^\al{}_\bt - \dr_\m K_\la{}^\al{}_\bt +
K_\la{}^\g{}_\bt K_\m{}^\al{}_\g - K_\m{}^\g{}_\bt K_\la{}^\al{}_\g
\ee
has the temporal component 
\beq
R_{\la\m}{}^0{}_\bt=0. \label{990}
\eeq
 Then the equation for
a Jacobi vector field $u$ along a geodesic $c$ reads
\beq
\dot q^\bt\dot q^\m(\nabla_\bt(\nabla_\m u^\al) - R_{\la\m}{}^\al{}_\bt u^\la)=
0,\qquad \nabla_\bt\dot q^\al=0, \label{991}
\eeq
where $\nabla_\m$ denote the covariant derivatives relative to the
connection $\wt K$ \cite{kob}. Due to the relation
(\ref{990}), the equation (\ref{991}) for the temporal
component $u^0$ of a Jakobi field takes the form
\be
\dot q^\bt\dot q^\m(\dr_\m\dr_\bt u^0 + K_\m{}^\g{}_\bt \dr_\g u^0)=0.
\ee
We chose its solution 
\beq
u^0=0 \label{993}
\eeq
because all non-relativistic geodesics obey the constraint $\dot q^0=0$.

Note that, in the case of a quadratic Lagrangian $L$, the equation (\ref{991})
coincides with the Jacobi equation
\be
u^jd_0(\dr_j\dot\dr_iL) + d_0(\dot u^j\dot\dr_i\dot\dr_j L) -u^j\dr_i\dr_jL=0
\ee
for a Jacobi field on solutions of the Lagrange equations for $L$. This
equation is the Lagrange equation for the vertical extension $L_V$ of the
Lagrangian $L$ \cite{giach99,book98,book99} (see also \cite{ditt}).

Let us consider a quadratic Newtonian system 
with a Riemannian mass tensor $m_{ij}$.  Given a reference frame
$(t,q^i)$, this mass tensor is extended to the Riemannian metric 
\beq
\ol g_{00}=1, \qquad \ol g_{0i}=0, \qquad \ol g_{ij}=m_{ij} \label{994}
\eeq
on $Q$.
However, its covariant derivative with respect to the connection $\wt K$
(\ref{cqg101}) does not vanish in general. 
Nevertheless, due to the relations (\ref{nl7}) and (\ref{993}), the
well-known formula 
\ben
&& \op\int_a^b(\ol g_{\la\m}(\dot q^\al\nabla_\al u^\la)( \dot q^\bt\nabla_\bt
u^\m) + R_{\la\m\al\nu} u^\la u^\bt \dot q^\m\dot q^\nu)
dt + \label{995}\\
&& \qquad \ol g_{\la\m}\dot q^\al\nabla_\al u^\la u'^\m\mid_{t=a} -
\ol g_{\la\m}\dot q^\al\nabla_\al u^\la u'^\m\mid_{t=b} =0 \nonumber
\een
for a
Jacobi vector field $u$ along a geodesic $c$ takes place. Accordingly, the
following assertions also remain true \cite{kob}.

\begin{prop} \label{1}
If the sectional curvature $R_{\la\m\al\nu} u^\la u^\bt \dot q^\m \dot
q^\nu$ is positive on a geodesic $c$, this geodesic has no conjugate
points.
\end{prop}

\begin{prop} \label{2}
If the sectional curvature $R_{\la\m\al\nu} u^\la u^\bt v^\m v^\nu$, where
$u$, $v$ are arbitrary unit vectors on a Riemannian manifold $Q$ less than
$k<0$, then, for every geodesic, the distance between two consecutive conjugate
points is at most $\pi/\sqrt{k}$.  
\end{prop}

For instance, let us consider a one-dimensional motion described by the
Lagrangian
\be
L=\frac12 (\dot q^1)^2 -\f(q^1),
\ee
where $\f$ is a potential. The corresponding Lagrange equation is equivalent
to the geodesic one on the 2-dimensional space $\bR^2$ with respect to the
connection $\wt K$ whose non-zero component is $\wt K_0{}^1{}_0=-\dr_1\f$.
The curvature of $\wt K$ has the non-zero component
\be
R_{10}{}^1{}_0=\dr_1\wt K_0{}^1{}_0=-\dr_1^2\f.
\ee
Choosing the Riemannian metric (\ref{994}) as
\be
\ol g_{11}=1, \quad \ol g_{01}=0, \quad \ol g_{00}=1,
\ee
we come to the formula (\ref{995}) 
\be
\op\int_a^b[(\dot q^\m\dr_\m u^1)^2- \dr_1^2\f (u^1)^2]dt=0.
\ee
for a Jacobi vector field $u$ which vanishes at points $a$ and $b$. Then
we obtain from Proposition
\ref{1} that, if $\dr_1^2\f<0$ at points of $c$, this motion has no
conjugate points.
In particular, let us consider the oscillator $\f= k(q^1)^2/2$.  In this case,
the sectional curvature is $R_{0101}=-k$, while the half-period of this
oscillator is exactly
$\pi/\sqrt{k}$ in accordance with Proposition \ref{2}.


\begin{thebibliography}{ederf}

\bibitem{cari93}  Cari\~nena J and Fern\'andez-N\'u\~nez J 1993 {\it Fortschr.
Phys.} {\bf 41} 517 

\bibitem{cramp} Crampin M, Mart\'{\i}nez E and Sarlet W 1996 {\it Ann.
Inst. H. Poincar\'e} {\bf 65A} 223.

\bibitem{ditt} Dittrich W and Reuter M 1992 {\it Classical and Quantum
Dynamics} (Springer: Berlin) 

\bibitem{eche}  Echeverr\'{\i}a Enr\'{\i}quez A,  Mu\~noz Lecanda  and
 Rom\'an Roy N 1991 {\it Rev. Math. Phys.} {\bf 3} 301 

\bibitem{eche95}  Echeverr\'{\i}a-Enr\'{\i}quez A,  Mu\~noz-Lecanda M and
 Rom\'an-Roy N 1995 {\it
J. Phys. A.} {\bf 28} 5553 
 
\bibitem{giach92} Giachetta G 1992
{\it J. Math. Phys.} {\bf 33} 1652

\bibitem{book} Giachetta G, Mangiarotti L and Sardanashvily G 1997 {\it New
Lagrangian and Hamiltonian Methods in Field Theory} (Singapore: World
Scientific)

\bibitem{giach99} Giachetta G, Mangiarotti L and Sardanashvily G 1999 {\it J.
Math. Phys.} {\bf 40} 1376.

\bibitem{kob} Kobayashi S and Nomizu K 1969 {\it Foundations of
Differential Geometry, V.II} ( N.Y.: Interscience Publishers)

\bibitem{kol} Kol\'a\v{r} I, Michor P and Slov\'ak J 1993 {\it Natural
Operations  in Differential Geometry} (Berlin: Springer-Verlag)

\bibitem{krupva} Krupkova O 1997 {\it The Geometry of Ordinary Variational
Equations} (Berlin: Springer-Verlag)

\bibitem{leon} De Le\'on M and Rodrigues P 1989 {\it Methods of
Differential Geometry in Analytical Mechanics} (Amsterdam: North-Holland)

\bibitem{book98} Mangiarotti L and Sardanashvily G 1998 {\it Gauge Mechanics}
(Singapore: World Scientific)

\bibitem{book99} Mangiarotti L, Obukhov Yu and Sardanashvily G 1999 {\it
Connections in Classical and Quantum Field Theory} (Singapore: World
Scientific)


\bibitem{massa} Massa E and Pagani E 1994 {\it Ann. Inst.
Henri Poincar\'e} {\bf 61} 17

\bibitem{mora} Morandi G, Ferrario C, Lo Vecchio G, Marmo G and Rubano C 1990
 {\it Phys. Rep.} {\bf 188} 147

\bibitem{sard98} Sardanashvily G 1998 
{\it J. Math. Phys.} {\bf 39} 2714


\bibitem{sau} Saunders D 1989 {\it The Geometry of Jet Bundles} 
(Cambridge: Cambr. Univ. Press)


\end{thebibliography}
\end{document}